\documentclass[pra,aps,showpacs,twocolumn]{revtex4}

\usepackage{amsmath,amsfonts,amssymb,caption,hyperref,color,epsfig,graphics,subfigure,graphicx,latexsym,mathrsfs,placeins,theorem,url,verbatim,epstopdf,float,adjustbox}

\usepackage{caption}
\usepackage{booktabs}
\usepackage{array}
\usepackage{afterpage}
\newcommand{\id}{\mathbb{I}}
\allowdisplaybreaks[3]

\hypersetup{colorlinks,linkcolor={blue},citecolor={blue},urlcolor={black},pdfborder={0 0 0}} 

\makeatletter

\newcommand{\Rmnum}[1]{\expandafter\@slowromancap\romannumeral #1@}
\makeatother
\newtheorem{definition}{Definition}
\newtheorem{proposition}[definition]{Proposition}
\newtheorem{Lemma}{Lemma}

\newtheorem{fact}{Fact}
\newtheorem{Theorem}[definition]{Theorem}

\newtheorem{conjecture}[definition]{Conjecture}

\newtheorem{remark}[definition]{Remark}

\newtheorem{example}{Example}
\newtheorem{question}[definition]{Question}

\newcommand{\Tr}{\mathrm{Tr}}

\def\squareforqed{\hbox{\rlap{$\sqcap$}$\sqcup$}}
\def\qed{\ifmmode\squareforqed\else{\unskip\nobreak\hfil
		\penalty50\hskip1em\null\nobreak\hfil\squareforqed
		\parfillskip=0pt\finalhyphendemerits=0\endgraf}\fi}
\def\endenv{\ifmmode\;\else{\unskip\nobreak\hfil
		\penalty50\hskip1em\null\nobreak\hfil\;
		\parfillskip=0pt\finalhyphendemerits=0\endgraf}\fi}
\newenvironment{proof}{\noindent \textbf{{Proof.~} }}{\qed}
\def\Dbar{\leavevmode\lower.6ex\hbox to 0pt
	{\hskip-.23ex\accent"16\hss}D}
\makeatletter
\def\url@leostyle{%
	\@ifundefined{selectfont}{\def\UrlFont{\sf}}{\def\UrlFont{\small\ttfamily}}}
\makeatother
\def\bcj{\begin{conjecture}}
	\def\ecj{\end{conjecture}}
\def\bcr{\begin{corollary}}
	\def\ecr{\end{corollary}}
\def\bd{\begin{definition}}
	\def\ed{\end{definition}}
\def\bea{\begin{eqnarray}}
\def\eea{\end{eqnarray}}
\def\bem{\begin{enumerate}}
	\def\eem{\end{enumerate}}
\def\bex{\begin{example}}
	\def\eex{\end{example}}
\def\bim{\begin{itemize}}
	\def\eim{\end{itemize}}
\def\bl{\begin{lemma}}
	\def\el{\end{lemma}}
\def\bma{\begin{bmatrix}}
	\def\ema{\end{bmatrix}}
\def\bpf{\begin{proof}}
	\def\epf{\end{proof}}
\def\bpp{\begin{proposition}}
	\def\epp{\end{proposition}}
\def\bqu{\begin{question}}
	\def\equ{\end{question}}
\def\br{\begin{remark}}
	\def\er{\end{remark}}
\def\bt{\begin{theorem}}
	\def\et{\end{theorem}}

\def\btb{\begin{tabular}}
	\def\etb{\end{tabular}}

\newcommand{\nc}{\newcommand}

\nc{\bbA}{\mathbb{A}} \nc{\bbB}{\mathbb{B}} \nc{\bbC}{\mathbb{C}}
\nc{\bbD}{\mathbb{D}} \nc{\bbE}{\mathbb{E}} \nc{\bbF}{\mathbb{F}}
\nc{\bbG}{\mathbb{G}} \nc{\bbH}{\mathbb{H}} \nc{\bbI}{\mathbb{I}}
\nc{\bbJ}{\mathbb{J}} \nc{\bbK}{\mathbb{K}} \nc{\bbL}{\mathbb{L}}
\nc{\bbM}{\mathbb{M}} \nc{\bbN}{\mathbb{N}} \nc{\bbO}{\mathbb{O}}
\nc{\bbP}{\mathbb{P}} \nc{\bbQ}{\mathbb{Q}} \nc{\bbR}{\mathbb{R}}
\nc{\bbS}{\mathbb{S}} \nc{\bbT}{\mathbb{T}} \nc{\bbU}{\mathbb{U}}
\nc{\bbV}{\mathbb{V}} \nc{\bbW}{\mathbb{W}} \nc{\bbX}{\mathbb{X}}
\nc{\bbZ}{\mathbb{Z}}

\nc{\bA}{{\bf A}} \nc{\bB}{{\bf B}} \nc{\bC}{{\bf C}}
\nc{\bD}{{\bf D}} \nc{\bE}{{\bf E}} \nc{\bF}{{\bf F}}
\nc{\bG}{{\bf G}} \nc{\bH}{{\bf H}} \nc{\bI}{{\bf I}}
\nc{\bJ}{{\bf J}} \nc{\bK}{{\bf K}} \nc{\bL}{{\bf L}}
\nc{\bM}{{\bf M}} \nc{\bN}{{\bf N}} \nc{\bO}{{\bf O}}
\nc{\bP}{{\bf P}} \nc{\bQ}{{\bf Q}} \nc{\bR}{{\bf R}}
\nc{\bS}{{\bf S}} \nc{\bT}{{\bf T}} \nc{\bU}{{\bf U}}
\nc{\bV}{{\bf V}} \nc{\bW}{{\bf W}} \nc{\bX}{{\bf X}}
\nc{\bZ}{{\bf Z}}

\nc{\cA}{{\cal A}} \nc{\cB}{{\cal B}} \nc{\cC}{{\cal C}}
\nc{\cD}{{\cal D}} \nc{\cE}{{\cal E}} \nc{\cF}{{\cal F}}
\nc{\cG}{{\cal G}} \nc{\cH}{{\cal H}} \nc{\cI}{{\cal I}}
\nc{\cJ}{{\cal J}} \nc{\cK}{{\cal K}} \nc{\cL}{{\cal L}}
\nc{\cM}{{\cal M}} \nc{\cN}{{\cal N}} \nc{\cO}{{\cal O}}
\nc{\cP}{{\cal P}} \nc{\cQ}{{\cal Q}} \nc{\cR}{{\cal R}}
\nc{\cS}{{\cal S}} \nc{\cT}{{\cal T}} \nc{\cU}{{\cal U}}
\nc{\cV}{{\cal V}} \nc{\cW}{{\cal W}} \nc{\cX}{{\cal X}}
\nc{\cZ}{{\cal Z}}

\nc{\hA}{{\hat{A}}} \nc{\hB}{{\hat{B}}} \nc{\hC}{{\hat{C}}}
\nc{\hD}{{\hat{D}}} \nc{\hE}{{\hat{E}}} \nc{\hF}{{\hat{F}}}
\nc{\hG}{{\hat{G}}} \nc{\hH}{{\hat{H}}} \nc{\hI}{{\hat{I}}}
\nc{\hJ}{{\hat{J}}} \nc{\hK}{{\hat{K}}} \nc{\hL}{{\hat{L}}}
\nc{\hM}{{\hat{M}}} \nc{\hN}{{\hat{N}}} \nc{\hO}{{\hat{O}}}
\nc{\hP}{{\hat{P}}} \nc{\hR}{{\hat{R}}} \nc{\hS}{{\hat{S}}}
\nc{\hT}{{\hat{T}}} \nc{\hU}{{\hat{U}}} \nc{\hV}{{\hat{V}}}
\nc{\hW}{{\hat{W}}} \nc{\hX}{{\hat{X}}} \nc{\hZ}{{\hat{Z}}}

\nc{\hn}{{\hat{n}}}

\def\max{\mathop{\rm max}}
\def\min{\mathop{\rm min}}

\newcommand{\bra}[1]{\langle#1|}
\newcommand{\ket}[1]{|#1\rangle}

\def\Dbar{\leavevmode\lower.6ex\hbox to 0pt
	{\hskip-.23ex\accent"16\hss}D}

\begin{document}
	\title{Quantifying Coherence and Genuine Multipartite Entanglement : A Framework Based on Witness Operators and Frobenius Norm Distance}
	
	\author{Mingyu Liu}\email[]
	{2024210568@buct.edu.cn}
	\affiliation{College of Information Science and Technology,
		Beijing University of Chemical Technology, Beijing 100029, China}
	\author{Xian Shi}\email[]
	{shixian01@gmail.com (corresponding author)}
	\affiliation{College of Information Science and Technology,
		Beijing University of Chemical Technology, Beijing 100029, China}

	\date{\today}
	\begin{abstract}
	Quantifying the entanglement and coherence of quantum systems is very important for theory and real-world applications. In this paper, we propose a method to evaluate lower bounds for several widely used coherence measures and genuine multipartite entanglement (GME) measures. Our approach is resource-efficient and computationally feasible. Finally, we present a practical framework for estimating these quantum resources in various physical scenarios.
	\end{abstract}

	\maketitle

  \section{Introduction}
  The quantum advantage  refers to the capability of quantum computers to surpass classical computers in performing certain tasks \cite{NielsenChuang2010,Harrow2017}. Quantum entanglement and quantum coherence (QC) are two crucial resources that establish this advantage \cite{RevModPhys.91.025001}. For instance, coherent superposition and entanglement provide the essential quantum phenomena that allow two famous algorithms, Shor’s algorithm and Grover’s algorithm \cite{Shor1995PolynomialTimeAF,10.1145/237814.237866}, to achieve their computational advantage. In some recent quantum application achievements, efforts have also been made to reduce the consumption of these two quantum resources. For example, a new quantum processor architecture was proposed, significantly enhancing qubit coherence time and operational precision \cite{Stassi2020}. Similarly, researchers have introduced a quantum error-correcting code that substantially reduces the entanglement resource consumption of physical qubits. \cite{PhysRevA.111.012444}. Given the critical role of these quantum resources, their precise and efficient quantification is crucial for advancing quantum technology.

  Quantum entanglement is one of the most famous non-classical phenomena in quantum mechanics, characterized by strong correlations between multiple particles, and it transcends the explanatory scope and fundamental assumptions of classical physics. For bipartite systems, entanglement is a well-established concept. Researchers have developed various measures for its quantification, such as the concurrence \cite{PhysRevA.54.3824,PhysRevLett.78.5022,PhysRevLett.80.2245,PhysRevA.64.052304,PhysRevA.64.042315,Badziag2001ConcurrenceIA}, the geometric measure of entanglement \cite{Barnum2001MonotonesAI,PhysRevA.68.042307,PhysRevA.70.022322}, and the negativity or extensions thereof \cite{HORODECKI19961,PhysRevLett.77.1413,PhysRevA.58.883,PhysRevA.65.032314,PhysRevA.68.062304}. Moving beyond bipartite systems, multipartite systems have introduced more intricate forms of entanglement, notably genuine multipartite entanglement (GME). While numerous criteria have existed for detecting multipartite entanglement \cite{Streltsov_2010,PhysRevLett.104.210501,PhysRevLett.106.020405}, the quantification of GME remains a significant and active research area, which is crucial for understanding its full utility as a quantum resource. This research specifically addresses the quantitative assessment of GME, contributing to its practical quantification by generalizing existing bipartite entanglement measures to multipartite scenarios. As entanglement is a pivotal resource, its accurate quantification is paramount, and QC represents another equally vital aspect of quantum mechanics that similarly demands precise quantification.

  \begin{figure*}[htbp]
  \centering
  \hspace{-1 cm} 
  \includegraphics[scale=0.14]{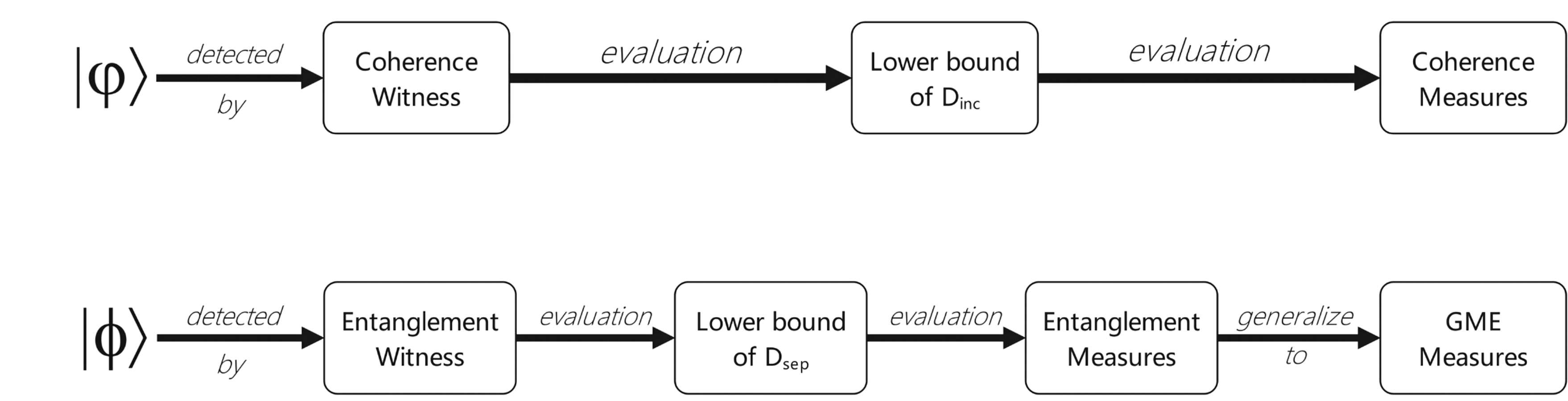}
  
  \caption{Evaluating Lower Bounds of Coherence Measures and GME Measures through Witness Operators. Where $D_{inc}$ and $D_{sep}$ are the minimum distances in terms of Frobenius norm between the given state and all incoherent or separable states, respectively.}
  \label{figure1}
  \end{figure*}

Alongside entanglement, QC stands as another indispensable quantum resource. Stemming from the fundamental principles of quantum mechanics, it manifests in the non-diagonal elements of a quantum state's density matrix when represented in a given basis. This property has enabled advancements significantly more efficient than classical methods, driving progress in diverse fields such as low-temperature thermodynamics \cite{Narasimhachar2015,PRXQuantum.3.040323,PhysRevResearch.5.043184}, quantum key distribution \cite{PhysRevA.99.062325,PhysRevA.109.012614,PhysRevLett.130.220801,PhysRevX.11.041016}, and quantum metrology \cite{PhysRevLett.96.010401,PhysRevA.99.022314,PhysRevA.94.010102}. With the rapid advancement of quantum information science, a surge of interest has emerged in methods to quantify coherence resources \cite{PhysRevLett.113.140401,RevModPhys.89.041003,PhysRevA.93.012334,PhysRevLett.119.150405,GUO2023106611,PhysRevA.107.022408}. Notably, Baumgratz and colleagues established a foundational resource theory framework for measuring coherence \cite{PhysRevLett.113.140401}, Streltsov and colleagues later demonstrated the explicit evaluation of the geometric measure of coherence \cite{PhysRevA.68.042307,Streltsov_2010}. While both entanglement and coherence are fundamentally important, fully characterizing them continued to present significant practical challenges, which we aim to address.

While entanglement and coherence are of fundamental importance, their precise quantification presents a major bottleneck. The standard method, full quantum state tomography, is resource-intensive and proves particularly challenging for large-qubit systems, especially when quantifying genuine multipartite entanglement (GME) \cite{PhysRevApplied.13.054022}. Consequently, determining how to quantify the coherence and GME of an unknown state with fewer resources has remained an intriguing problem. To address this challenge, this research introduces a novel method for evaluating the lower bounds of coherence and GME measures, as conceptually illustrated in Figure \ref{figure1}. The approach significantly reduces quantum resource consumption by strategically employing generic witness operators in conjunction with a distance metric based on the Frobenius norm ($F$-norm). The strateg has fundamentally avoided the need for comprehensive quantum state tomography, offering a computationally feasible and experimentally practical pathway for estimating these crucial quantum resources. In our previous work \cite{shi2024lower}, we established a framework utilizing witness operators and the Frobenius norm distance to derive lower bounds for bipartite entanglement measures. In this manuscript, we significantly extend this framework in two key directions: first, by adapting the methodology to quantify quantum coherence, and second, by generalizing the method from bipartite entanglement to the more complex realm of genuine multipartite entanglement (GME). To demonstrate the framework's utility, especially its integration with machine learning, we require a realistic simulation of a quantum system's evolution.

For the training, we have modeled system decoherence using the Lindblad master equation \cite{Balslev1968,PhysRevA.111.042626,PhysRevResearch.4.023216} in conjunction with amplitude damping operators. The Lindblad master equation is a powerful and widely accepted tool for describing open quantum systems; its application has enabled the accurate modeling of non-unitary dynamics that arise from system-environment interactions. This is crucial for simulating quantum systems realistically. Amplitude damping operators \cite{PhysRevA.67.064301}, in turn, are particularly adept at capturing energy dissipation processes, which represent a primary cause of decoherence in many quantum systems. Together, the Lindblad master equation and amplitude damping operators have provided a realistic and computationally efficient framework for simulating the decoherence of quantum states due to environmental influences. This robust simulation approach has ensured the reliability of the coherence measure calculations and the subsequent neural network training presented herein.

In this work, we establish several key results. First, we derive effective lower bounds for three crucial coherence measures: the $l_1$-norm measure \cite{PhysRevLett.113.140401}, the relative entropy of coherence \cite{PhysRevLett.113.140401}, and the geometric measure of coherence \cite{PhysRevA.68.042307,Streltsov_2010}. Second, we demonstrate a method that generalizes the lower bounds of bipartite entanglement measures, which serves as a foundational step. We leverage this method to establish lower bounds for GME measures, as exemplified by a calculation targeting the concurrence of a GME state. Finally, to demonstrate the framework's broad utility, we not only validate its robustness and applicability through comprehensive simulation studies, but also integrate machine learning techniques—specifically by training a neural network as a classifier—to perform efficient, low-cost coherence probing that determines if the system's coherence falls below a predefined threshold. The model training results under the master equation show that the neural network successfully learned the mapping from the witness expectation value to the coherence measures.

The remainder of this paper is organized as follows: In Section II, we provide the preliminary knowledge necessary to understand our proposed methodology. In Section III, we elaborate on our main theoretical results concerning the derived lower bounds. Then, in Section IV, we present the simulation calculations that validate the method. In Section V, we explore a potential practical application of the method—its integration with machine learning. Finally, in Section VI, we conclude this work, summarizing the main findings and outlining future directions.

  \section{Preliminary knowledge}\label{section2}
  In this section, we first give the mathematical definition of entanglement and coherence, and define two distances based on the $F$-norm. These distances will play crucial roles in deriving lower bounds for entanglement and coherence measures. Then we introduce three commonly used coherence measures and three commonly used entanglement measures for bipartite states. Finally, we will provide a method for generalizing bipartite measures to GME measures.
  \subsection{The definition of entanglement and coherence}
   The mathematical definition of bipartite entanglement is as follows \cite{RevModPhys.81.865}: In a $d$-dimensional Hilbert space $\mathcal{H}$, a bipartite mixed state $\rho$ is considered separable if it can be expressed in the form:
  \begin{align}
      \rho=\sum^d_ip_i\rho_i^A\otimes\rho_i^B,
  \end{align}
  where $p_i\geq0$ and $\sum_ip_i=1$, with $\rho_A$ and $\rho_B$ being the density matrices of subsystems $A\in\mathcal{H}$ and $B\in\mathcal{H}$, respectively. If a mixed state $\rho$ cannot be represented in the above form, it is entangled.

  The mathematical definition of QC is as follows:
  In a $d$-dimensional Hilbert space $\mathcal{H}$, by selecting a basis $\{|i\rangle\}$, the quantum state $\rho$ is defined as incoherent with respect to the basis $\{\ket{i}\}$ if its density matrix representation in this basis is diagonal. This means the density matrix can be expressed as:
 $$\rho = \sum_{i=1}^d p_i \ket{i}\bra{i},$$
 where $p_i\ge 0$ and $\sum_{i=1}^d p_i = 1$. In this case, all off-diagonal elements are zero, i.e., for all $i \neq j$, $\rho_{ij} = \langle i|\rho|j \rangle= 0$.
 Otherwise, the state $\rho$ is said to exhibit quantum coherence with respect to the basis $\{\ket{i}\}$.

\subsection{The distances based on the $F$-norm}
Let's recall the $F$-norm of an $m \times n$ matrix $A$, denoted as $||A||_F$, can be defined by the sum of the squares of its elements as:
$$
||A||_F = \sqrt{\sum_{i=1}^m \sum_{j=1}^n |A_{ij}|^2}
$$
where $A_{ij}$ represents the element in the $i$-th row and $j$-th column of matrix $A$. Then we define the distance between the mixed state $\rho_{AB}$ and the set of separable states in term of $\|\cdot\|_F$ as:
\begin{align}
	D_{sep}(\rho)=\min_{\sigma\in S}\|\rho-\sigma\|_F,
\end{align}
where the minimum takes over all the separable states $\sigma\in Sep(A:B)$. 
Similarly, we define the distance between the mixed state $\rho$ and the set of incoherent states in term of $\|\cdot\|_F$ as:
\begin{align}
	D_{inc}(\rho)=\min_{\sigma\in \mathcal{I}}\|\rho-\sigma\|_F.\label{Dinc1}
\end{align}
where the minimum takes over all the incoherent states $\sigma\in \mathcal{I}$. 

 \begin{fact}\label{fact1}
 Since the $F$-norm is defined as $\|A\|_F = \sqrt{\sum_{i,j} |a_{ij}|^2}$, and $\sigma$ belongs to the incoherent state, it can only be a diagonal matrix. Therefore, the $\sigma$ closest to $\rho$ can only be the diagonal matrix $\delta$ composed of the diagonal elements of $\rho$, ensuring that all diagonal elements result zero after the subtraction operation. That is, 
 \begin{align}
    D_{inc}(\rho) = \|\rho - \delta\|_F.
 \end{align}
\end{fact}

  \subsection{Commonly used entanglement measures}\label{preB}
  Let's recall three commonly used entanglement measures. Assume $|\psi\rangle_{AB}$ is a bipartite pure state. Due to the Schmidt decomposition, $|\psi\rangle_{AB}$ can always be written as
  \begin{align}
      |\psi\rangle_{AB} = \sum_{i} \sqrt{\lambda_i} |i\rangle_A |i\rangle_B,
  \end{align}
where $\lambda_i \geq 0$, $\sum_{i} \lambda_i = 1$, and $\{|i\rangle_{A(B)}\}$ is an orthonormal basis of the Hilbert space $A(B)$. The entanglement of formation (EoF) of $|\psi\rangle_{AB}$ is given by
\begin{align}
E_f(|\psi\rangle_{AB}) = S(\rho_A) = - \sum_i \lambda_i \log_2 \lambda_i,
\end{align}
where $\lambda_i$ are the eigenvalues of $\rho_A = \operatorname{Tr}_B |\psi\rangle_{AB} \langle \psi|$. For a mixed state $\rho_{AB}$, and EoF for a mixed state $\rho_{AB}$ is defined by the convex roof extension method,
\begin{align}
E_f(\rho_{AB}) = \min_{\{p_i,|\phi_i\rangle_{AB}\}} \sum_i p_i E_f(|\phi_i\rangle_{AB}),
\end{align}
where the minimum takes over all the decompositions of $\rho_{AB} = \sum_i p_i |\phi_i\rangle_{AB} \langle \phi_i|$, with $p_i \geq 0$ and $\sum_i p_i = 1$.
The other important entanglement measure is the concurrence (C). The concurrence of a pure state $|\psi\rangle_{AB}$ is defined as \cite{PhysRevLett.80.2245,PhysRevA.64.042315}
\begin{align}
C(|\psi\rangle_{AB}) = \sqrt{2(1 - \operatorname{Tr} \rho_A^2)} = \sqrt{2(1 - \sum_i \lambda_i^2)}.
\end{align}
For a mixed state $\rho_{AB}$, it is defined as
\begin{align}
C(\rho_{AB}) = \min_{\{p_i,|\phi_i\rangle_{AB}\}} \sum_i p_i C(|\phi_i\rangle_{AB}),
\end{align}
where the minimum takes over all the decompositions of $\rho_{AB} = \sum_i p_i |\phi_i\rangle_{AB} \langle \phi_i|$ with $p_i \geq 0$ and $\sum_i p_i = 1$. Another entanglement measure is the geometrical entanglement measure $E_g$. For a pure state $|\psi\rangle_{AB} = \sum_{i=0}^{d-1} \sqrt{\lambda_i} |ii\rangle$,
\begin{align}
E_g(|\psi\rangle_{AB}) = \min_{\sigma = |\phi_1\rangle_A |\phi_2\rangle_B} 1 - |\langle \psi|\sigma|\psi\rangle| = 1 - \lambda_0.
\end{align}
where $\lambda_0$ is the largest Schmidt coefficient, and the minimum takes over all the product state $\sigma$.
For a mixed state $\rho_{AB}$, it is defined as
\begin{align}
E_g(\rho_{AB}) = \min_{\{p_i,|\phi_i\rangle_{AB}\}} \sum_i p_i E_g(|\phi_i\rangle_{AB}),
\end{align}
where the minimum takes over all the decompositions of $\rho_{AB} = \sum_i p_i |\phi_i\rangle_{AB} \langle \phi_i|$ with $p_i \geq 0$ and $\sum_i p_i = 1$.
 
 \subsection{Commonly used coherence measures}
 Next, the specific mathematical expressions for the three commonly used coherence measures introduced earlier are provided: the $l_1$-norm coherence, the relative entropy coherence, and the geometric measure of coherence. For arbitrary matrix, we define that $\| \mathbf{A} \|_{\text{asum}} = \sum_{i=1}^{m} \sum_{j=1}^{n} |a_{ij}|$.
 Assuming $\rho$  is the density matrix of the quantum state to be measured and $\mathcal{I}$ represents the set of all incoherent states, the $l_1$-norm measure of coherence, which can be denoted as \cite{PhysRevLett.113.140401}
  \begin{align}
	C_{l_1}(\rho) = \min_{\delta \in \mathcal{I}} \|\rho - \delta\|_{\text{asum}} = \sum_{i \neq j} |\rho_{i,j}|.\label{Cl1}
  \end{align}
  where the minimum is achieved when $\delta$ equals the completely dephased $\rho$, i.e., the diagonal matrix consisting only of the diagonal elements of $\rho$. The relative entropy coherence, which can be denoted as \cite{PhysRevLett.113.140401}
  \begin{align}
	C_{ref}(\rho) = S(\rho_{diag}) - S(\rho),\label{Ccof}
  \end{align}
  where $S(\cdot)$ means von Neumann entropy, $\rho_{diag}$ is a diagonal matrix composed of diagonal elements of $\rho$. For a pure state $|\psi\rangle$, its coherence of formation is defined as $C_{cof}(|\psi\rangle)=S((|\psi\rangle\langle\psi|)_{diag})$ \cite{PhysRevApplied.13.054022}, which is equivalent to $C_{ref}$. Since the former applies to mixed states via the convex roof construction \cite{PhysRevLett.116.120404,PhysRevA.92.022124}, we will primarily use $C_{cof}$ in the following discussion.
  According to \cite{PhysRevLett.115.020403}, the geometric measure of coherence can be denoted as:
  \begin{align}
      C_g(\rho)=1-\max_{\delta\in \mathcal{I}}F(\rho,\delta).\label{Cg}
  \end{align} 
   It is important to note that when it comes to QC, we tend to view the system as a whole \cite{PhysRevLett.113.140401,RevModPhys.89.041003,HU20181,RevModPhys.91.025001}. Therefore, in the following coherence measures, subscripts representing subsystems are not used, to emphasize this point. 
   
  \subsection{Definition of GME measures}\label{preD}
  A pure $N$-partite quantum state $|\psi\rangle$ is said to exhibit GME if it cannot be expressed as a tensor product under any nontrivial bipartition, i.e.,
\begin{equation*}
|\psi\rangle \neq |\psi_A\rangle \otimes |\psi_{\overline{A}}\rangle, \quad \forall \text{ bipartitions } A|\overline{A}.
\end{equation*}
For a mixed state $\rho$, it is genuinely multipartite entangled if it cannot be written as a convex mixture of states separable across some bipartition:
\begin{equation*}
\rho = \sum_i p_i \rho_A^i \otimes \rho_{\overline{A}}^i, \quad \forall \text{ bipartitions } A|\overline{A},
\end{equation*}
where $p_i$ is a probability distribution, and $\rho_A^i$ and $\rho_{\overline{A}}^i$ are density matrices of subsystems $A$ and $\overline{A}$, respectively.

  One approach to quantifying GME is to generalize measures of bipartite entanglement to multipartite systems.  
  Considering a bipartite entanglement measure $E$, one can define a GME measure as
\begin{align}
	E_{GME}(|\psi\rangle)=\min_{A|\overline{A}}E_{A|\overline{A}}(|\psi\rangle),
\end{align}
for an $N$-partite pure state $|\psi\rangle$, where $A$ represents all possible bipartitions $A|\overline{A}$ of $\{1,2,\cdots,N\}$. For instance, suppose that $i_1,i_2,\cdots,i_N$ is an arbitrary order of $1,2,\cdots,N$. The subset A contains $\{i_1,\cdots,i_k\}$ and $\overline{A}$ = $\{i_{k+1},\cdots,i_N\}$ with $1\leq k\leq N-1$, since $A$ and $\overline{A}$ are two nonempty subsets.

For any GME measure $E_{GME}(\cdot)$ applied to an arbitrary $N$-partite state $\rho$, we have
\begin{align}
	E_{GME}(\rho)&=\inf_{\{p_i,|\psi_i\rangle\}}\sum_ip_iE_{GME}(|\psi_i\rangle)\label{EGME1}\\
	&= \inf_{\{p_i,|\psi_i\rangle\}}\sum_ip_i\min_{A|\overline{A}}E_{A|\overline{A}}(|\psi_i\rangle)\label{EGME2},
\end{align}
where $A|\overline{A}$ denotes an arbitrary bipartition, the infimum takes over all the decompositions of $\rho_{A\overline{A}}=\sum_ip_i|\psi_i\rangle_{A\overline{A}}\langle\psi_i|$, with $p_i\geq0$ and $\sum_ip_i=1$, and the minimum is taken over all possible bipartitions of the N-partite system. $E_{A|\overline{A}}$ means the corresponding bipartite entanglement measure between subsystem $A$ and $\overline{A}$.
  \section{Main results}
\subsection{Lower bounds of coherent quantifiers based on coherent witness}
  
Following the reference \cite{shi2024lower}, we can utilize the dual expression of the $F$-norm and normalize the general coherence witness to have a $F$-norm of 1. This approach has allowed for the derivation of the lower bound of the $D_{inc}$ measure for mixed states with respect to general coherence witnesses. 

    Assume $W_0$ is a generic coherence witness of $\mathcal{H}_d$, let $ a = \frac{\operatorname{Tr} W_0}{d},b = \sqrt{\operatorname{Tr}(W_0^\dagger W_0) - \frac{(\operatorname{Tr} W_0)^2}{d}}, \quad W_1 = \frac{W_0 - a I \otimes I}{b}. $ Based on the property of $\|\cdot\|_F$, we have
    \begin{align*}
    D_{inc}(\rho)&=\min_{\sigma\in \mathcal{I}}\max_{\|W\|_F=1}\operatorname{Tr}[W(\rho-\sigma)]\\
    &\geq |\operatorname{Tr}[W_1(\rho-\delta)]|\\
    &=|\operatorname{Tr}[\frac{W_0}{b}(\rho-\delta)]-\frac{a}{b}\operatorname{Tr}[\rho-\delta]|\\
    &=|\operatorname{Tr}[\frac{W_0}{b}(\rho-\delta)]|\\
    &=|\frac{1}{b}[\operatorname{Tr}(W_0\rho)-\operatorname{Tr}(W_0\delta)]|\\
    &\geq-\frac{1}{b}\operatorname{Tr}(W_0\rho),
\end{align*}
where the first equation is the dual norm of the $F$-norm, and $\delta$ is a diagonal matrix with diagonal elements identical to $\rho$. The last inequality holds because $\operatorname{Tr}(W_0\delta)\geq 0$.

\begin{Theorem}\label{theorem_pure}
Assume $|\psi\rangle$ is a pure state $|\psi\rangle=\sum_{i=0}^{d-1}\lambda_i|i\rangle$, satisfying the normalization condition $\sum_{i=0}^{d-1}|\lambda_i|^2=1$. Let $I$ be the set of incoherent states (diagonal density matrices in the basis $\{|i\rangle\}$), then 
\begin{align}
	D_{inc}(|\psi\rangle)&=\min_{\sigma\in I}\| |\psi\rangle\langle \psi|-\sigma \|_F\label{minDinc}\\
	&=\sqrt{1-\sum_{i=0}^{d-1}|\lambda_i|^4}.
\end{align}
\end{Theorem}
\begin{Theorem}\label{theorem_mix}
Assume $\rho$ is a mixed state, then
\begin{align}
	C_{l_1}(\rho)&\geq D_{inc}(\rho),\\
	C_g(\rho)&\geq \frac{1}{4}D_{inc}^2(\rho),\\
	C_{cof}(\rho)&\geq -log_2(1-D_{inc}^2(\rho)).\label{cof}
\end{align}
\end{Theorem}
Theorem \ref{theorem_pure} presents $D_{inc}$ as a $F$-norm-based measure of coherence for pure states, which quantifies the quantum superposition components in a pure state that cannot be described by classical probability under a specific basis. This measure provides a direct method to compute coherence through state vector coefficients, which is convenient for both theoretical analysis and numerical calculations. We leverage this property in the proof of Theorem \ref{theorem_mix} (Equation \ref{cof}).
Theorem \ref{theorem_mix} establishes three conservative lower bounds for commonly-used coherence measures based on $D_{inc}$. As shown previously, $D_{inc}$ itself has a lower bound derived from general coherence witnesses $W_0$. Therefore, Theorem \ref{theorem_mix} serves as a critical link that enables us to estimate the coherence strength of unknown quantum states using general coherence witnesses. See Appendix \ref{Appendix2} for the proof of the above theorems.
\begin{figure*}[htbp]
  \centering
  \hspace{0 cm} 
  \includegraphics[scale=0.1]{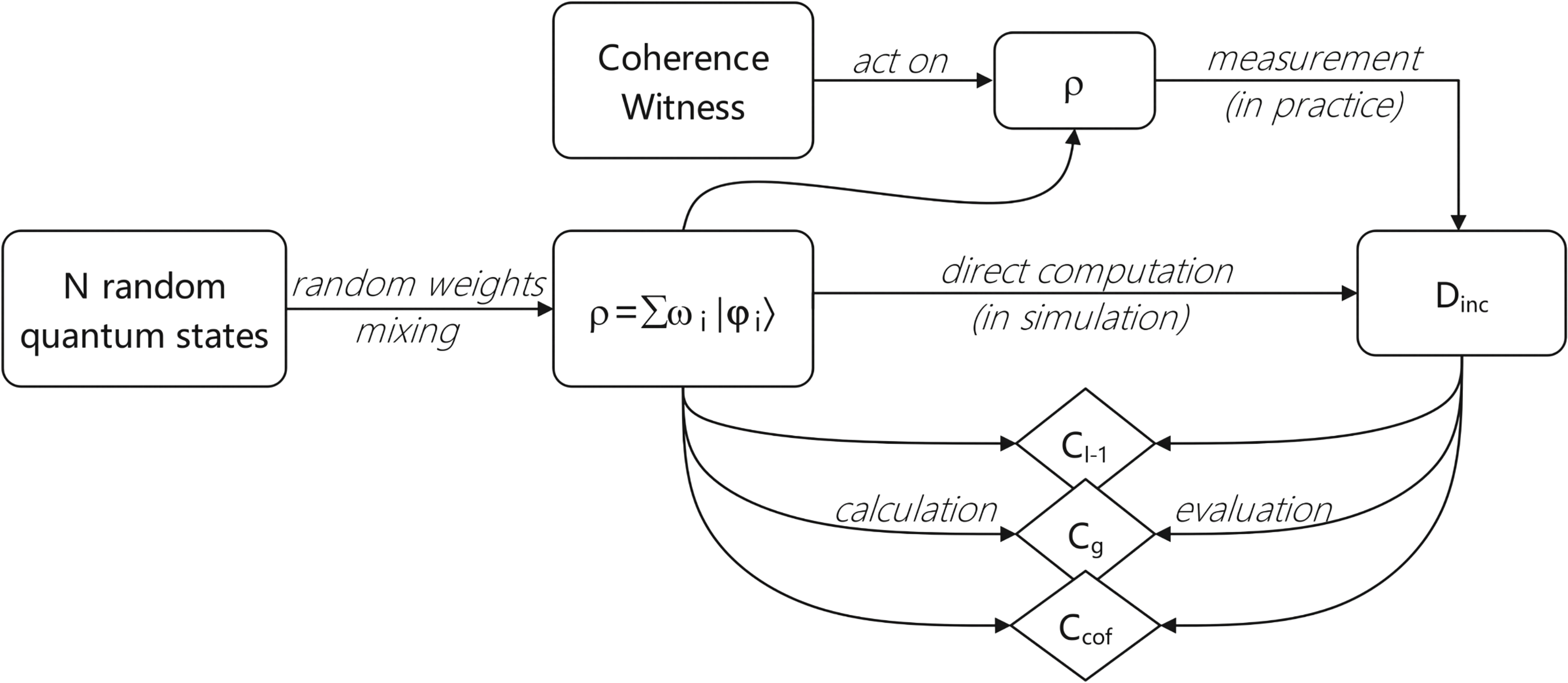}

  \caption{Simulation of lower bounds for three coherence measures.}
  \label{figure3}
\end{figure*}
  \subsection{Lower bounds of GME measures based on entanglement witness}

  In this section, the bipartite entanglement measure, concurrence (introduced in Section \ref{preB}), is first generalized to GME concurrence using the method described in Section \ref{preD}. A legitimate and effective generic GME witness is then employed to evaluate its lower bound. Using this process as an example, a general form is then proposed for evaluating GME measures based on generic GME witnesses.
  
  Before we begin, we first need to supplement the method (see Lemma \ref{lemma1}) proposed by the authors in \cite{shi2024lower} for calculating the lower bound of $D_{sep}$ in bipartite states, making it applicable to subsystems with different dimensions, i.e., for any bipartition of GME:  

  Assume $W_0$ is a generic GME witness of $\mathcal{H}_{d}$. For any $m, n \in \mathbb{Z}^+$ such that $m \otimes n = d$, the Hilbert space under an arbitrary bipartition $A|\overline{A}$ can be represented as: $\mathcal{H}_d^{(A|\overline{A})}=\mathcal{H}_m\otimes\mathcal{H}_n$, let $a = \frac{\mathrm{Tr} W_0}{mn}, \quad b = \sqrt{\mathrm{Tr}(W_0^\dagger W_0) - \frac{(\mathrm{Tr} W_0)^2}{mn}}, \quad W_1 = \frac{W_0 - a I_m \otimes I_n}{b}.$ Based on the property of $\|\cdot\|_F$, we have
  \begin{align*}
  D^{(A|\overline{A})}_{\text{sep}}(\rho_{AB}) &= \min_{\sigma \in Sep(A:B)} \max_{\|W\|_F=1} \mathrm{Tr}(W_1 (\rho_{AB} - \sigma_{AB})) \\
  &\geq |\mathrm{Tr} W_1 (\rho - \omega)| \\
  &= \left| \mathrm{Tr} \left( \frac{W_0}{b} (\rho - \omega) - \frac{a}{b} \mathrm{Tr} (\rho - \omega) \right) \right| \\
  &= \left| \mathrm{Tr} \left[ \frac{W_0}{b} (\rho - \omega) \right] \right| \\
  &\geq -\frac{1}{b} \mathrm{Tr}( W_0 \rho),
  \end{align*}
in the first inequality, $\omega$ is the optimal separable state, $W_1$ is a Hermitian operator and $\|W_1\|_F = 1$. The last inequality is due to $\mathrm{Tr} W_0 \omega \geq 0$. For different $A|\overline{A}$, the tensor product of the dimensions of the two subsystems always satisfies $mn = d$, hence the parameter $b$ remains the same. In fact, what differs is $\omega_{A|\overline{A}}$, but it is discarded in the final step, just as explained for the last inequality in the preceding context. Therefore, the result obtained here is the most conservative lower bound, that is:
\begin{align}
    \min_{A|\overline{A}}D_{\text{sep}}^{(A|\overline{A})}(\rho_{AB})\geq-\frac{1}{b} \mathrm{Tr} (W_0 \rho).\label{minDsep}
\end{align}

  \subsubsection{Lower bounds of GME concurrence}
  
Assume that $\rho = \sum_i p_i |\psi_i\rangle\langle\psi_i|$ is the optimal decomposition for $N$-partite state $\rho$ to achieve the infimum of $C_{GME}(\rho)=\inf_{\{p_i,|\psi_i\rangle\}}\sum_ip_iC_{GME}(|\psi_i\rangle)$. According to \cite{shi2024lower}, we have $C_{BE}(\rho)\geq \sqrt{2}D_{sep}(\rho)$, where $BE$ represent bipartite entanglement. For all possible bipartitions $A|\overline{A}$, we can calculate the concurrence of $\rho$:
  \begin{align}
      C_{BE}^{(A|\overline{A})}(\rho)&=\sum_ip_iC_{BE}^{(A|\overline{A})}(|\psi_i\rangle)\\
      &\geq\sum_ip_i\sqrt{2}D_{sep}(|\psi_i\rangle)\\
      &\geq\sqrt{2}D_{sep}(\rho)\label{CBE1}.
  \end{align}
As shown in Eq.(\ref{minDsep}), $D_{sep}$ is the most conservative lower bound and is independent of the choice of bipartitions. Then, we have
  \begin{align}
      C_{GME}(\rho)&=\sum_ip_iC_{GME}(|\psi_i\rangle)\\
      &=\sum_ip_i\min_{A|\overline{A}}C_{BE}^{(A|\overline{A})}(|\psi_i\rangle)\\
      &\geq\sum_ip_i\sqrt{2}D_{sep}(|\psi_i\rangle)\\
      &\geq\sqrt{2}D_{sep}(\rho)\\
      &\geq-\frac{\sqrt{2}}{b}Tr(W_0\rho)\label{CGME1},
  \end{align}
 where $W_0$ is a generic GME witness on $\mathcal{H}_{d}$, and is valid for $\rho$, that is, if $\rho$ is a GME state, then $Tr(W_0\rho)<0$. $b=\sqrt{Tr(W_0^\dagger W_0)-\frac{(TrW_0)^2}{d}}$. The first and second equations follow from Eq.(\ref{EGME2}).
  \subsubsection{The general form}
  Using the evaluation of the lower bound for GME concurrence as an illustrative example, we propose a general form for evaluating various GME measures by leveraging entanglement witnesses. Based on the analysis of several bipartite entanglement measures \cite{shi2024lower,PhysRevApplied.13.054022}, our proposed method is applicable to, but not limited to, GME-extended measures such as the entanglement of formation, the concurrence, the geometric entanglement measure, and the G-concurrence.

\begin{Theorem}\label{GME-entanglement}
   Let $D_{\text{sep}}(\rho)$ be the $F$-norm distance between the given $N$-partite state $\rho$ and the nearest separable state, and let 
\begin{multline}
    \mathcal{A} = \{f \mid f:\mathbb{R} \to [0,1], \\ 
    f \text{ is monotonically increasing and convex}\}
\end{multline}
For any bipartite state $\sigma$ and a given entanglement measure $E(\cdot)$ such that there exists $f \in \mathcal{A}$, $E(\sigma) \geq f(D_{\text{sep}}(\sigma))$ then the lower bound of the GME measure generalized by $E(\cdot)$ can be expressed as:
\begin{align}
    E_{GME}(\rho)&\geq f\{-\frac{1}{b}Tr(W_0\rho)\}\label{GME_final},
\end{align}
where, where $W_0$ is a generic GME witness on $\mathcal{H}_{d}$, and is valid for $\rho$, $b=\sqrt{Tr(W_0^\dagger W_0)-\frac{(TrW_0)^2}{d}}$.
\end{Theorem} 

\bpf
Assume $\rho$ is an arbitrary $N$-partite state, and $\sigma$ is an arbitrary bipartite state. When $D_{\text{sep}}(\rho)=0$, $\rho$ is a fully separable $N$-partite state.
For such a state, the infimum in Eq.~(\ref{EGME1}) is achieved by a decomposition of $\rho$ into $N$-partite product states, i.e., $\rho = \sum_i p_i |\psi_i^{(\text{inf})}\rangle\langle\psi_i^{(\text{inf})}|$.
For these product states $|\psi_i^{(\text{inf})}\rangle$, any bipartite entanglement $E_{A|\overline{A}}(|\psi_i^{(\text{inf})}\rangle)$ is zero for any bipartition $A|\overline{A}$.
Consequently, by Eq.~(\ref{EGME2}):
$$ E_{GME}(|\psi_i^{(\text{inf})}\rangle) = \min_{A|\overline{A}}E_{A|\overline{A}}(|\psi_i^{(\text{inf})}\rangle) = 0. $$
Thus, from Eq.~(\ref{EGME1}), it follows that:
\begin{align*}
    E_{GME}(\rho) &= \sum_i p_i E_{GME}(|\psi_i^{(\text{inf})}\rangle) \\
                  &= \sum_i p_i \cdot 0 \\
                  &= 0.
\end{align*}
When $D_{sep}>0$ and $E(\sigma)\geq f(D_{sep}(\sigma))$, we have:
\begin{align*}
    E_{GME}(\rho)&=\sum_ip_iE_{GME}(|\psi_i\rangle)\\
    &= \sum_ip_i\min_{A|\overline{A}}E^{(A|\overline{A})}(|\psi_i\rangle)\\
    &\geq \sum_ip_if\{\min_{A|\overline{A}}D_{sep}^{(A|\overline{A})}(|\psi_i\rangle)\}  \\
    &\geq f\{\min_{A|\overline{A}}D_{sep}^{(A|\overline{A})}(\rho)\}\\
    &\geq f\{-\frac{1}{b}Tr(W_0\rho)\},
\end{align*}
The first inequality holds because $f(\cdot)$ is a monotonically increasing function, and since $f(\cdot)$ is also a convex function, the second inequality follows from Jensen's inequality \cite{Jensen1906}. The  final equality is due to Eq.(\ref{minDsep}).

\epf

  \section{Numerical Verification}

  In this part, we verify the validity of the aforementioned theoretical framework through numerical simulations. We evaluate the lower bounds for the coherence measures of various quantum states and the entanglement measures of two typical three-qubit GME states, respectively.

First, we performed numerical simulations on the coherence of multipartite mixed states to verify the relationships between the three coherence measures ($C_{l_1}$, $C_g$, and $C_{cof}$) proposed in Theorem \ref{theorem_mix} and their respective lower bounds. For each test case, we began by generating a set of $N$ random pure states $\ket{\psi_i}$ in a $d=4$ Hilbert space, which corresponds to a two-qubit system. These pure states were obtained by applying random unitary operators, drawn from the Circular Unitary Ensemble (CUE), to a reference state, such as $\ket{00}$. Subsequently, these pure states were mixed with equal weights ($p_i = 1/N$) to form a mixed state $\rho = \sum_{i=1}^{N} p_i \ket{\psi_i}\bra{\psi_i}$.

To directly test the validity of the inequalities in Theorem \ref{theorem_mix}, we calculated the exact value of $D_{inc}(\rho)$ for each generated mixed state $\rho$. This was achieved by directly computing the F-norm of the off-diagonal elements of the density matrix through Fact \ref{fact1}. At the same time, the exact values of the coherence measures $C_{l_1}$,$C_g$, and $C_{cof}$ were also calculated for each state $\rho$.

The results are shown in Table \ref{tab:1}. This table directly compares the exact values of the coherence measures (columns 1, 3, and 5) with the corresponding theoretical bounds calculated using the exact value of $D_{inc}$(columns 2, 4, and 6). The data clearly demonstrate that the relationships established in Theorem \ref{theorem_mix} hold.  Since any witness-based estimation of $D_{inc}$ would yield a lower bound that is by definition less than or equal to its exact value, this direct verification of Theorem \ref{theorem_mix} also indirectly confirms the overall validity of our framework for estimating coherence.

  \begin{table}[htbp]
	\centering
	\caption{Three coherence measures and their corresponding lower bounds}
	\label{tab:1}  
	\begin{tabular}{cccc ccc}
		\hline\hline\noalign{\smallskip}	
		NMPS & $C_{l_1}$ & $D_{inc}$ & $C_g$ & $\frac{1}{4}D_{inc}^2$ & $C_{cof}$ & $-log_2(1-D_{inc}^2)$  \\
		\noalign{\smallskip}\hline\noalign{\smallskip}
		$2_a$ & 1.796 & 0.569 & 0.370 & 0.081 & 0.959 & 0.392\\
		$2_b$ & 1.554 & 0.595 & 0.450 & 0.088 & 0.995 & 0.437\\
		$2_c$ & 2.167 & 0.656 & 0.498 & 0.108 & 1.217 & 0.563\\
		$5_a$ & 0.886 & 0.283 & 0.137 & 0.020 & 0.335 & 0.083\\
		$5_b$ & 0.765 & 0.237 & 0.083 & 0.014 & 0.236 & 0.058\\
		$5_c$ & 1.434 & 0.448 & 0.243 & 0.050 & 0.634 & 0.223\\
		$10_a$ & 0.646 & 0.211 & 0.047 & 0.011 & 0.134 & 0.046\\
		$10_b$ & 1.066 & 0.336 & 0.106 & 0.028 & 0.312 & 0.120\\
		$10_c$ & 0.740 & 0.245 & 0.067 & 0.015 & 0.184 & 0.062\\
		$20_a$ & 0.613 & 0.212 & 0.04435 & 0.01120 & 0.125 & 0.046\\
		$20_b$ & 0.734 & 0.236 & 0.055 & 0.014 & 0.161 & 0.057\\
		$20_c$ & 0.471 & 0.153 & 0.027 & 0.006 & 0.076 & 0.024\\
		$100$ & 0.324 & 0.107 & 0.01165 & 0.00284 & 0.033 & 0.011\\
	\noalign{\smallskip}\hline
	\end{tabular}
        \\[10pt] 
	\footnotesize 
	\textit{Note:} NMPS stands for "number of mixed pure states," i.e., the number of pure states contributing to the mixed state. In the table, columns 1, 3, and 5 are exact values, while columns 2, 4, and 6 are lower bound values.
\end{table}

Next, the methods were demonstrated for two GME scenarios. We then calculated the lower bounds of three GME measures generalized from \ref{preB} for the 3-qubit GHZ state and 3-qubit W state \cite{Greenberger1989}.

For GME states, a commonly used entanglement witness is: $W_0 = c\id - \ket{\psi}\bra{\psi}$ (where $\ket{\psi}$ is the pure GME state we are considering, $P_\psi = \ket{\psi}\bra{\psi}$, and $c$ is a real constant chosen such that $W_0$ is a valid GME witness and $\Tr(W_0 P_\psi) < 0$), the parameter $b^2$ simplifies significantly.
Given $\Tr(W_0) = cd - 1$ and $\Tr(W_0^2) = \Tr((c\id - P_\psi)^2) = \Tr(c^2\id - 2cP_\psi + P_\psi^2) = c^2d - (2c-1)$,
\begin{align*}
    b^2 &= \left(c^2 d - 2c + 1\right) - \frac{(cd-1)^2}{d} \\
        &= c^2 d - 2c + 1 - \frac{c^2 d^2 - 2cd + 1}{d} \\
        &= c^2 d - 2c + 1 - (c^2 d - 2c + 1/d) \\
        &= 1 - \frac{1}{d} = \frac{d-1}{d}
\end{align*}
Thus, $b = \sqrt{\frac{d-1}{d}}$. The expectation value $\Tr(W_0 P_\psi) = c \Tr(P_\psi) - \Tr(P_\psi^2) = c-1$.
The lower bound becomes:
\begin{equation} \label{eq:dsep_simplified_bound}
    D_{\text{sep}}(P_\psi) \ge -\frac{c-1}{\sqrt{(d-1)/d}} = (1-c)\sqrt{\frac{d}{d-1}}
\end{equation}

\noindent The $3$-qubit GHZ state is defined as:
\begin{equation}\label{3qGHZ}
    \ket{\text{GHZ}_3} = \frac{1}{\sqrt{2}}(\ket{0}^{\otimes 3} + \ket{1}^{\otimes 3})
\end{equation}
Let $\rho_{\text{GHZ}_3} = \ket{\text{GHZ}_3}\bra{\text{GHZ}_3}$. The total dimension is $d=2^3=8$. Existing research \cite{GUHNE20091,TERHAL2000319} shows that by choosing c=1/2, $W_0$ can serve as an effective GME witness for GHZ states.
Using Equation \eqref{3qGHZ}:
\begin{equation}
    D_{\text{sep}}(\rho_{\text{GHZ}_3}) \ge \left(1-\frac{1}{2}\right)\sqrt{\frac{8}{8-1}} \approx0.5345\label{GHZ-Dsep}
\end{equation}
In \cite{shi2024lower}, we learn that for bipartite states, the three entanglement measures introduced in \ref{preB} and $D_{sep}$ satisfy the following relationship:
\begin{Theorem}\label{shi}
Assume $\rho_{AB}$ is a bipartite mixed state, then
\begin{align*}
C(\rho_{AB}) &\geq \sqrt{2} D_{\text{sep}}(\rho_{AB}), \\
E_f(\rho_{AB}) &\geq -\log_2(1 - D_{\text{sep}}^2(\rho_{AB})), \\
E_g(\rho_{AB}) &\geq D_{\text{sep}}^2(\rho_{AB}).
\end{align*}
\end{Theorem}
Theorem \ref{shi} supplements the specific expression of the mapping relationships -- $f$, in Theorem \ref{GME-entanglement}. Therefore, combining Eqs. (\ref{GME_final}) and (\ref{GHZ-Dsep}), we have
\begin{align*}
C(\rho_{\text{GHZ}_3}) &\geq 0.7560 \\
E_f(\rho_{\text{GHZ}_3}) &\geq 0.4854 \\
E_g(\rho_{\text{GHZ}_3}) &\geq 0.2857.
\end{align*}

\noindent The 3-qubit W state is defined as:
\begin{equation}
    \ket{W_3} = \frac{1}{\sqrt{3}}(\ket{100} + \ket{010} + \ket{001})
\end{equation}
Let $\rho_{\text{W}_3} = \ket{\text{W}_3}\bra{\text{W}_3}$. The total dimension of the Hilbert space is $d=2^3=8$.
For constructing a GME witness of the form $W_0 = c\id - \rho_{\text{W}_3}$, a suitable choice for $N$-qubit W states is $c=(N-1)/N$ \cite{GUHNE20091}. For $N=3$, this yields $c=(3-1)/3 = 2/3$. Using Equation (\ref{eq:dsep_simplified_bound}) for $\rho_{\text{W}_3}$ with $c=2/3$ and $d=8$:
\begin{equation}
\begin{split}
    D_{\text{sep}}(\rho_{\text{W}_3}) &\ge \left(1-\frac{2}{3}\right)\sqrt{\frac{8}{8-1}}\approx 0.3563 \label{W3-Dsep_continuation}
\end{split}
\end{equation}
Therefore, combining Eqs. (\ref{GME_final}) and (\ref{W3-Dsep_continuation}), we have
\begin{align*}
C(\rho_{\text{W}_3}) &\geq 0.5039 \\
E_f(\rho_{\text{W}_3}) &\geq 0.1958 \\
E_g(\rho_{\text{W}_3}) &\geq 0.1270.
\end{align*}

In practice, the parameter $b$ is determined entirely by the witness operator $W_0$. Therefore, estimating the lower bound for $D_{sep}$ only requires measuring the expectation value $Tr(W_0\rho)$.

\section{Application: Real-Time Coherence Monitoring via Machine Learning}

\begin{figure*}[t]
    
    \centering
    \includegraphics[width=0.8\textwidth]{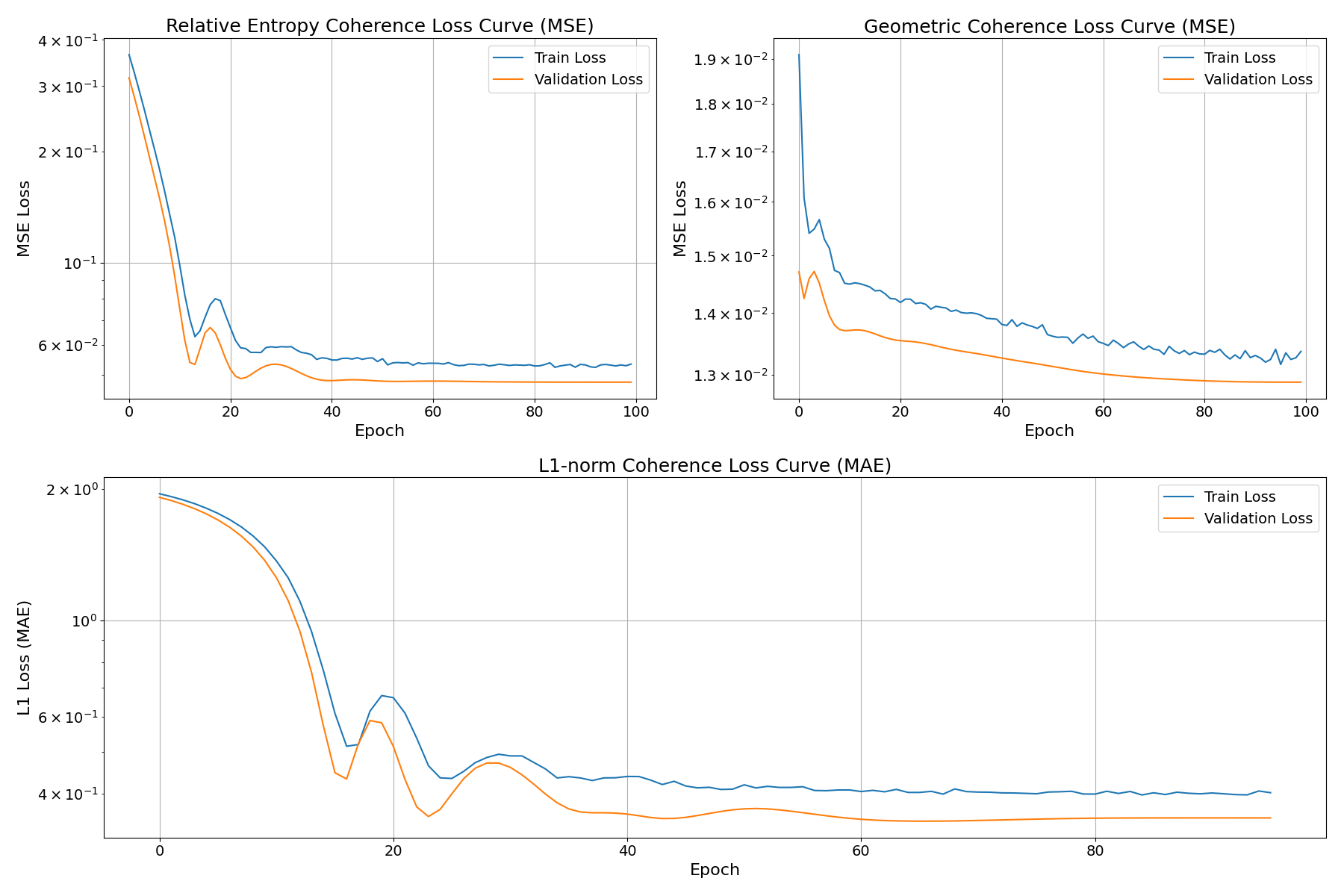}
    \caption{Loss curves of the regression task}
    \label{fig1}
    
\end{figure*}

\begin{figure*}[t]
    \centering
    \includegraphics[width=0.8\textwidth]{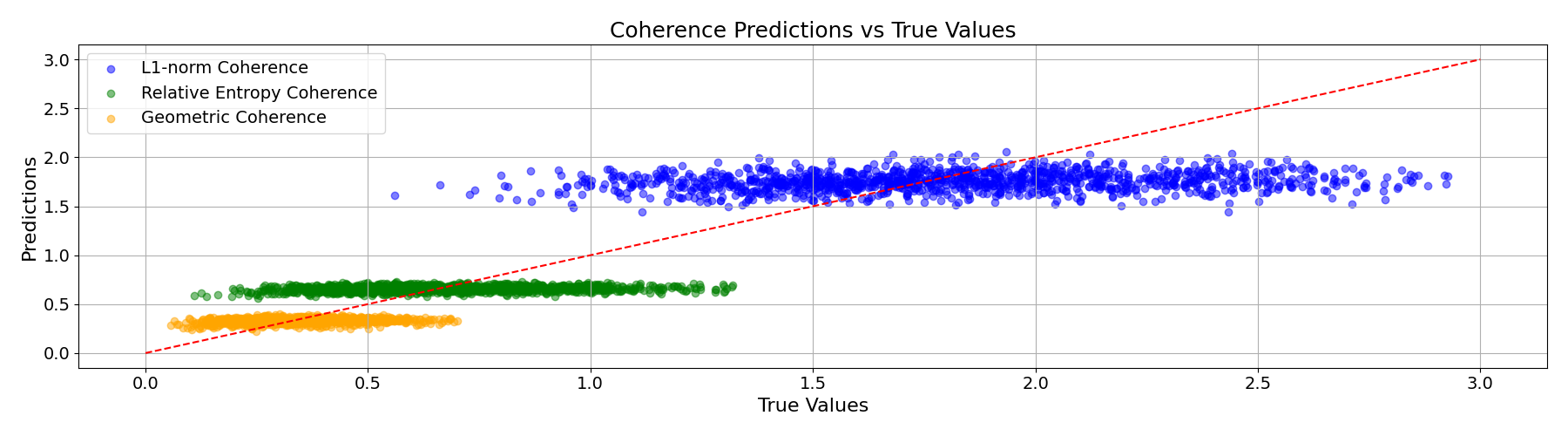}
    \caption{Scatter plots of the true values against the predicted values for the coherence measures}
    \label{fig2}
\end{figure*}

In the preceding sections, we established a general theoretical framework for estimating lower bounds of quantum resources via witness operators. However, the efficacy of this framework for a \textit{completely unknown state}—one with no prior information—is contingent on finding a suitable generic witness, which remains a formidable challenge.

In this section, we pivot to explore the framework's application in a different, yet highly practical scenario: \textbf{the real-time monitoring of a quantum system whose initial state $\rho_0$ is known, as it undergoes an unknown dynamical evolution}. This task is distinct from quantifying a completely unknown state; here, we are concerned with an \textit{unknown evolved state}, $\rho(t) = \mathcal{E}(\rho_0)$, where the evolution $\mathcal{E}$ is the unknown element. Such monitoring is crucial in many experimental contexts, such as assessing the quality of quantum gates or evaluating noise in a quantum channel.

This premise allows us to move beyond generic, state-independent witnesses. Since the initial state $\rho_0$ is known, we can perform a one-time calibration by constructing an optimal, state-dependent witness operator tailored specifically to it. Following the strategy for creating a phase-matched probe, we identify the off-diagonal element of $\rho_0$ with the largest magnitude, say $\rho_{0,ij}$, and define the witness as $W_0 = e^{-i\phi}|i\rangle\langle j| + e^{i\phi}|j\rangle\langle i|$, where $\phi = \arg(\rho_{0,ij})$. This operator is optimally aligned to probe the dominant coherence term in the initial state. Even with this optimized witness, the relationship between its expectation value on the evolved state, $\text{Tr}(W_0\rho(t))$, and the true coherence measures of $\rho(t)$ remains complex and non-linear. Therefore, we employ a machine learning model to learn this sophisticated mapping, effectively calibrating the witness measurement to provide a high-fidelity estimation of the true coherence.

\subsection{Experimental Design}

We simulate a 2-qubit system undergoing decoherence via local amplitude damping channels to generate a dataset of 8000 samples. The system's evolution is governed by the Lindblad master equation \cite{Lindblad1976}, with the Hamiltonian set to $H=0$ to focus purely on dissipative effects:
\begin{align}
\frac{d\rho(t)}{dt} = &-i[H, \rho(t)] + \sum_{k=1}^N \left( L_k \rho(t) L_k^\dagger - \frac{1}{2} \left\{ L_k^\dagger L_k, \rho(t) \right\} \right),\nonumber\\
L_k =& \sqrt{\gamma} |0\rangle_k\langle 1|,
\end{align}
where $H$ is the system's Hamiltonian, $\{ \cdot, \cdot \}$ denotes the anticommutator, $|0\rangle_k\langle 1|$ is the lowering operator, and $\gamma$ represents the energy decay rate.

For each sample, an initial mixed state $\rho_0$ is generated, and its corresponding phase-matched witness operator $W_0$ is constructed according to the aforementioned method. This fixed witness is then used to measure the system after decoherence, yielding an expectation value $m(t) = \text{Tr}(W_0\rho(t))$. The input features for the neural network are a 33-dimensional vector, comprising the flattened real and imaginary parts of $\rho_0$ and the scalar value $m(t)$. This simulates a practical scenario where a system is thoroughly characterized once to establish an optimal measurement basis, which is then used for low-cost, real-time monitoring.

Two machine learning tasks were designed to evaluate the framework's utility:
\begin{enumerate}
    \item \textbf{Regression Task:} A neural network is trained to predict the precise values of three coherence measures ($C_{l_{1}}$, $C_{g}$, and $C_{cof}$). The loss function for $C_{l_{1}}$ is Mean Absolute Error (MAE), while Mean Squared Error (MSE) is used for $C_{g}$ and $C_{cof}$.
    \item \textbf{Classification Task:} The model is trained as a binary classifier to detect if the system's coherence has fallen below a predefined threshold, defined as a fraction ($x$) of its initial value.
\end{enumerate}

For specific details on the implementation and parameters, please refer to the Appendix \ref{sec:reproducibility}.

\subsection{Results and Analysis}

The trained models were evaluated on an unseen test set. For the \textbf{regression task}, as shown in Fig.\ref{fig1}, the model demonstrated exceptional performance in predicting $C_{g}$ and $C_{cof}$, achieving an MSE below 0.05 for both. This indicates a high-fidelity capture of these coherence features. In contrast, the prediction for $C_{l_1}$ was less precise, yielding an MAE of 0.3562. This discrepancy, which is visually evident in Fig.\ref{fig2}, can be explained by the underlying physics of our feature selection. The witness operator $W_0$ is constructed to be phase-matched to the off-diagonal element of $\rho_0$ with the largest magnitude. Consequently, the measured expectation value $\text{Tr}(W_0\rho(t))$ primarily tracks the evolution of this single, dominant coherence pathway. $C_g$ and $C_{cof}$, being related to fidelity and entropy respectively, are global properties of the state that are significantly influenced by such dominant terms. In contrast, $C_{l_1}$ is the sum of the absolute values of \textit{all} off-diagonal elements. Information from a single, targeted witness is likely insufficient to reconstruct this total sum, especially when other, smaller coherence terms evolve differently. The input features, therefore, do not contain enough information for the neural network to accurately learn the mapping for $C_{l_1}$. This limitation of the chosen feature set is a key insight that should be acknowledged.

For the \textbf{classification task}, the model proved effective at identifying the coherence threshold, as shown in Fig.\ref{fig1}. For threshold factors between 0.5 and 0.6, the model achieved over 80\% accuracy for $C_{cof}$ and $C_{g}$, demonstrating practical utility in moderately decoherent scenarios. The differing accuracy trend for $C_{l_1}$ is attributed to its statistical properties and potential dataset imbalances at varying thresholds.

In summary, the integration of phase-matched witness operators—calibrated using initial state information—with machine learning presents a powerful and resource-efficient paradigm for the dynamic monitoring of quantum systems.

\section{Conclusion}
\label{sec:conclusion}
This paper has presented and thoroughly validated a novel framework designed to quantify fundamental quantum resources, QC and GME, by establishing their lower bounds. The methodology centrally employs the strategic use of witness operators combined with the $F$-norm distance. This combination has significantly improved the practicality of assessing these resources, which are critical for the realization of quantum information tasks.

The theoretical contributions have included the derivation of effective lower bounds for several important classes of coherence measures. Furthermore, this approach has been successfully generalized to address the complex challenge of GME quantification in multipartite systems. The robustness and applicability of this framework have been confirmed through comprehensive simulation studies across various quantum state configurations, thereby demonstrating its utility for reliable resource estimation.

Beyond its theoretical foundations, the practical utility of the framework has been investigated, particularly by interfacing witness-based measurements with machine learning techniques. Neural networks have been employed to refine the estimation of quantum resources. These networks have learned to predict exact coherence values from witness expectation values and to effectively probe system coherence levels against predefined thresholds. This integration has pointed towards promising methods for the real-time characterization and monitoring of dynamic quantum systems.

A principal advantage of the proposed methodology has been its significant gains in resource efficiency and computational feasibility. When compared to conventional techniques, such as full quantum state tomography, which often compare extensive measurements and complex post-processing, this approach has provided a substantially simpler, less resource-intensive, and more experimentally feasible path to quantum resource quantification. This enhanced accessibility is a key benefit for experimentalists.

The implications of this research have extended beyond coherence and entanglement, offering a universal conceptual foundation for developing similar lower-bound estimation techniques for other crucial quantum properties. Future investigations will aim to further refine the precision of these bounds, adapt the methodology to diverse physical systems, and enhance collaborations with advanced machine learning models. 

\section*{Acknowledgements}
This work was supported by the National Natural Science Foundation of China (Grant No. 12301580).

  \appendix

  \section{Proofs of Theorems}\label{Appendix2}
  
  \begin{Lemma}\label{lemma1}\cite{shi2024lower} 
  Assume $W_0$ is a generic entanglement witness of $\mathcal{H}_d \otimes \mathcal{H}_d$, let $a = \frac{\mathrm{Tr} W_0}{d^2}, \quad b = \sqrt{\mathrm{Tr}(W_0^\dagger W_0) - \frac{(\mathrm{Tr} W_0)^2}{d^2}}, \quad W_1 = \frac{W_0 - a I \otimes I}{b}.$ Based on the property of $\|\cdot\|_F$, we have
  \begin{align*}
  D_{\text{sep}}(\rho_{AB}) &= \min_{\sigma \in Sep(A:B)} \max_{\|W_1\|_F=1} \mathrm{Tr}(W_1 (\rho_{AB} - \sigma_{AB})) \\
  &\geq |\mathrm{Tr} W_1 (\rho - \omega)| \\
  &= \left| \mathrm{Tr} \left( \frac{W_0}{b} (\rho - \omega) - \frac{a}{b} \mathrm{Tr} (\rho - \omega) \right) \right| \\
  &= \left| \mathrm{Tr} \left[ \frac{W_0}{b} (\rho - \omega) \right] \right| \\
  &\geq -\frac{1}{b} \mathrm{Tr} W_0 \rho,
  \end{align*}
in the first inequality, $\omega$ is the optimal separable state, $W_1$ is a Hermitian operator and $\|W_1\|_F = 1$. The last inequality is due to $\mathrm{Tr} W_0 \omega \geq 0$.
\end{Lemma}

\begin{Lemma}\label{lemma2}
Consider the square of the sum of the absolute values of all elements in the matrix $A\in \mathbb{R}^{m\times n}$
\begin{align}
    \left( \sum_{i=1}^{m} \sum_{j=1}^{n} |a_{ij}| \right)^2 = \sum_{i=1}^{m} \sum_{j=1}^{n} |a_{ij}|^2 + 2 \sum_{i<k} \sum_{j<l} |a_{ij}| \cdot |a_{kl}|.
\end{align}
Since all cross-product terms $|a_{ij}| \cdot |a_{kl}|$ are non-negative, we have:
\begin{align}
    \left( \sum_{i=1}^{m} \sum_{j=1}^{n} |a_{ij}| \right)^2 &\geq \sum_{i=1}^{m} \sum_{j=1}^{n} |a_{ij}|^2\\
    \sum_{i=1}^{m} \sum_{j=1}^{n} |a_{ij}| &\geq \sqrt{\sum_{i=1}^{m} \sum_{j=1}^{n} |a_{ij}|^2} \\
     \| \mathbf{A} \|_{\text{asum}}&\geq \| \mathbf{A} \|_F.
\end{align}
This proves that the $F$-norm of a matrix is less than or equal to the sum of the absolute values of all its elements.
\end{Lemma}
 
\noindent\textbf{Theorem \ref{theorem_pure} } \textit{Assume $|\psi\rangle$ is a pure state $|\psi\rangle=\sum_{i=0}^{d-1}\lambda_i|i\rangle$, satisfying the normalization condition $\sum_{i=0}^{d-1}|\lambda_i|^2=1$. Let $I$ be the set of incoherent states (diagonal density matrices in the basis $\{|i\rangle\}$). Then, the Frobenius distance to the set of incoherent states is given by}
\begin{align}
	D_{inc}(|\psi\rangle)&=\min_{\sigma\in I}\| |\psi\rangle\langle \psi|-\sigma \|_F\label{minDinc}\\
	&=\sqrt{1-\sum_{i=0}^{d-1}|\lambda_i|^4}.
\end{align}

\noindent\textit{Proof } Assume $|\psi\rangle = \sum_{i=0}^{d-1}\lambda_i|i\rangle$ is a pure state, and $\rho=|\psi\rangle\langle\psi|$. The elements of the density matrix $\rho$ are given by $\rho_{ij} = \lambda_i \lambda_j^*$. Thus, we have
\begin{equation}
	\rho =
	\begin{bmatrix}
|\lambda_0|^2 & \lambda_0 \lambda_1^* & \cdots & \lambda_0 \lambda_{d-1}^* \\
\lambda_1 \lambda_0^* & |\lambda_1|^2 & \cdots & \lambda_1 \lambda_{d-1}^* \\
\vdots & \vdots & \ddots & \vdots \\
\lambda_{d-1} \lambda_0^* & \lambda_{d-1} \lambda_1^* & \cdots & |\lambda_{d-1}|^2 \\
\end{bmatrix}.
	\label{eq:density_matrix_psi}
\end{equation}
From Fact \ref{fact1}, it is known that $\sigma=\rho_{\text{diag}}$, where $\rho_{\text{diag}}$ is a diagonal matrix composed of the diagonal elements of matrix $\rho$. Since each matrix element in the calculation of the $F$-norm results in a value that is greater than or equal to zero, $\sigma$ is constructed as a diagonal matrix composed of the diagonal elements of $\rho$. So
\begin{align}
	D_{inc}(|\psi\rangle) &= \sqrt{\sum_{i \neq j} |\lambda_i|^2 |\lambda_j|^2} \nonumber \\
	&= \sqrt{2\sum_{0\leq i<j \leq d-1}|\lambda_i|^2|\lambda_j|^2} \label{eq:proof_sum_offdiag_abs} \\
	&= \sqrt{\left(\sum_{i=0}^{d-1}|\lambda_i|^2\right)^2 - \sum_{i=0}^{d-1}(|\lambda_i|^2)^2} \label{eq:proof_identity_abs} \\
	&= \sqrt{\left(\sum_{i=0}^{d-1}|\lambda_i|^2\right)^2 - \sum_{i=0}^{d-1}|\lambda_i|^4} \label{eq:proof_power_abs} \\
	&= \sqrt{1 - \sum_{i=0}^{d-1}|\lambda_i|^4}. \label{eq:proof_final_result_abs}
\end{align}
The final equality (\ref{eq:proof_final_result_abs}) holds because $\rho$ is the density matrix of a normalized pure state, so its trace is $\text{Tr}(\rho) = \sum_{i=0}^{d-1} \rho_{ii} = \sum_{i=0}^{d-1} |\lambda_i|^2 = 1$.

\noindent\textbf{Theorem \ref{theorem_mix} }\textit{ Assume $\rho$ is a mixed state, then}
\begin{align}
	C_{l_1}(\rho)&\geq D_{inc}(\rho),\\
	C_g(\rho)&\geq \frac{1}{4}D_{inc}^2(\rho),\\
	C_{cof}(\rho)&\geq -log_2(1-D_{inc}^2(\rho)).
\end{align}

\noindent\textit{Proof} Assume $\rho$ is a mixed state, $\{p_i,|\psi_i\rangle\}$ is the optimal decomposition of $\rho$ in terms of $C_{l_1}(\rho)$,
\begin{align*}
	C_{l_1}(\rho)&=\inf_{\{p_i,|\psi_i\rangle\}}\sum_ip_iC_{l_1}(|\psi_i\rangle)\\
	&=\sum_ip_i\||\psi_i\rangle\langle\psi_i|-\sigma\|_{\text{asum}}\\
	&\geq\sum_ip_i\||\psi_i\rangle\langle\psi_i|-\sigma\|_F\\
	&=\sum_ip_iD_{inc}(|\psi_i\rangle)\\
	&\geq D_{inc}(\rho).
\end{align*}
The first inequality is valid according to Lemma \ref{lemma2}. 

Next assume $\rho$ is a mixed state, $\{p_i,|\psi_i\rangle\}$ is the optimal decomposition of $\rho$ in terms of $C_g(\rho)$,
\begin{align*}
	Cg(\rho)&=\sum_ip_iC_g(|\psi_i\rangle)\\
	&=\sum_ip_i[1-\max_{j\in I}|\langle j|\psi_i\rangle|^2]\\
	&=\sum_ip_i[1-F(|\psi_i\rangle,\sigma_i)]\\
	&\geq\sum_ip_iD^2(|\psi_i\rangle,\sigma_i)\\
	&=\frac{1}{4}\sum_ip_i\||\psi_i\rangle\langle\psi_i|-\sigma_i\|_{1}^2\\
	&\geq\frac{1}{4}\sum_ip_i\||\psi_i\rangle\langle\psi_i|-\sigma_i\|_F^2\\
	&=\frac{1}{4}\sum_ip_iD_{inc}^2(|\psi_i\rangle)\\
	&\geq\frac{1}{4}D_{inc}^2(\rho),
\end{align*}
where $\sigma_i$ denotes the optimal incoherent state chosen individually for each pure state component $|\psi_i\rangle$, $D(\cdot)$ means trace distance, $F(\cdot)$ means fidelity and $I$ is the set of all incoherent states. As $\|M^p\|^{\frac{1}{p}}$ for positive $M$ is monotone decreasing, the second inequality is valid. 

Finally assume $\rho$ is a mixed state, $\{p_i,|\psi_i\rangle\}$ is the optimal decomposition of $\rho$ in terms of $C_{cof}(\rho)$, $\sigma^{(i)}=|\psi_i\rangle\langle\psi_i|$
\begin{align*}
    C_{cof}(\rho)&=\sum_ip_iC_{cof}(|\psi_i\rangle)\\
    &=\sum_ip_i[S(\sigma^{(i)}_{diag})-S(\sigma^{(i)})]\\
    &=\sum_ip_iS(\sigma^{(i)}_{diag})\\ 
    &=-\sum_ip_i\sum_{j=0}^{d-1}\sigma^{(i)}_{jj}log_2\sigma^{(i)}{jj}\\
    &\geq-\sum_ip_ilog_2(\sum_{j=0}^{d-1}\sigma^{(i)}{jj}^2)\\
    &=-\sum_ip_ilog_2(\sum_{j=0}^{d-1}\lambda_j^4)_{(i)}\\
    &=-\sum_ip_ilog_2[(\sum_{j=0}^{d-1}\lambda_j^2)^2-2\sum_{0\leq j<k \leq d-1}\lambda_j^2\lambda_k^2]_{(i)}\\
    &=-\sum_ip_ilog_2(1-D_{inc}^2(|\psi_i\rangle)\\
    &\geq-log_2(1-D_{inc}^2(\rho)),
\end{align*}
where $\sigma_{diag}$ means the diagonal matrix composed of the diagonal elements of $\sigma$, $\sigma_{jj}$ means the $j^{th}$ diagonal element of $\sigma_{diag}$, $\lambda_i$ means the diagonal elements of $\sigma^{(i)}$. Since $\sigma^{(i)}$ is a pure state, therefor $S(\sigma^{(i)})=0$, the third equality is valid. According to matrix (\ref{eq:density_matrix_psi}) we have $\sigma_{jj}=\lambda_j^2$, so the fifth equality is valid. The first inequality follows from Jensen's inequality\cite{Jensen1906} for concave functions.

\section{Reproducibility Details}\label{sec:reproducibility}

This section provides a detailed account of the parameters and procedures required to reproduce the machine learning experiment. The implementation uses a computationally efficient equivalent to the continuous time evolution described by the master equation. The two methods are linked by the relationship between the channel's probability parameter $p$, the physical decay rate $\gamma$, and the evolution time $t$:
\begin{equation}
p = 1 - e^{-\gamma t}
\end{equation}
This formula shows that applying a quantum channel with a given parameter $p$ is equivalent to solving the master equation for a specific time duration $t$. A full derivation and theoretical background for this equivalence can be found in Chapter 8, Section 8.4 of "Quantum Computation and Quantum Information" by Nielsen and Chuang. Therefore, the code's method of using a random parameter $p$ for each sample is a practical shortcut to generate a diverse dataset, effectively sampling the results of the continuous Lindblad evolution at various random time points.

\subsubsection{Data Generation and Quantum Simulation}
A dataset of 8000 independent samples was generated. Each sample corresponds to a unique quantum state undergoing a single decoherence event. The parameters are detailed in Table \ref{tab:data_gen}.

\subsubsection{Feature Engineering and Witness Operator}
The input features for the model are constructed from both the initial state and the witness measurement after decoherence.

\begin{itemize}
    \item \textbf{Witness Operator ($W_0$):} A unique witness operator is constructed for each initial state $\rho_0$. The operator is designed to be phase-matched to the off-diagonal element of $\rho_0$ that has the largest absolute value.
    \item \textbf{Model Input Features:} The input for the neural network is a 33-dimensional feature vector for each sample. This vector is a concatenation of the following:
        \begin{itemize}
            \item The flattened real part of the initial density matrix $\rho_0$ (16 features).
            \item The flattened imaginary part of the initial density matrix $\rho_0$ (16 features).
            \item The real part of the witness expectation value, $\text{Tr}(W_0 \rho_t)$, where $\rho_t$ is the state after damping (1 feature).
        \end{itemize}
    \item \textbf{Data Scaling:} Before being fed to the model, the 33-dimensional input features are standardized using \texttt{sklearn.preprocessing.StandardScaler}.
\end{itemize}

\subsubsection{Neural Network Architecture: CoherenceNet}
The experiment utilizes a deep neural network with residual connections.
\begin{itemize}
    \item \textbf{Feature Extractor (\texttt{features} block)}:
    \begin{itemize}
        \item Fully-Connected Layer (33 $\to$ 128) with GELU activation, followed by LayerNorm.
        \item A \texttt{ResidualBlock} containing two fully-connected layers (128 $\to$ 128) and LeakyReLU activation.
        \item Dropout with a rate of $p=0.3$.
        \item Fully-Connected Layer (128 $\to$ 64) with LeakyReLU (alpha=0.01) activation.
        \item Dropout with a rate of $p=0.3$.
        \item Fully-Connected Layer (64 $\to$ 32) with SiLU (Sigmoid Linear Unit) activation.
    \end{itemize}
    \item \textbf{Output Head (\texttt{head} block)}:
    \begin{itemize}
        \item Fully-Connected Layer (32 $\to$ 16) with LeakyReLU (alpha=0.01) activation.
        \item A final Fully-Connected Layer (16 $\to$ 1) with a linear activation for regression tasks.
    \end{itemize}
\end{itemize}

\subsubsection{Training Protocol and Task Parameters}
The models for regression and classification tasks are trained using a shared set of core protocols, with task-specific configurations as detailed in Table \ref{tab:regression} and Table \ref{tab:classification}. Common settings include:
\begin{itemize}
    \item \textbf{Data Split:} The 8000 samples are split into 70\% for training (5600 samples), 15\% for validation (1200 samples), and 15\% for testing (1200 samples).
    \item \textbf{Optimizer:}Adam optimizer (\texttt{torch.optim.Adam}).
    \item \textbf{Hyperparameter Search:} A grid search is performed over the learning rate and weight decay.
    \item \textbf{Training Epochs:} A maximum of 100 epochs, with an early stopping patience of 30.
    \item \textbf{Learning Rate Scheduler:} \texttt{CosineAnnealingLR} is used to adjust the learning rate during training.
\end{itemize}

\begin{table*}[htbp]
\centering
\caption{Parameters for Training Data Generation}
\label{tab:data_gen}
\begin{tabular}{>{\raggedright}p{0.4\linewidth} >{\raggedright\arraybackslash}p{0.5\linewidth}}
\hline\hline\noalign{\smallskip}
\textbf{Parameter} & \textbf{Value / Method} \\
\noalign{\smallskip}\hline\noalign{\smallskip}
System Type & 2-qubit \\
Hilbert Space Dimension ($d$) & 4 \\
Dataset Size & 8000 samples \\
Initial Mixed State $\rho_0$ & Randomly weighted mixture of 3 random pure states \\
Initial State for Evolution & Pure State \\
Decoherence Model & PennyLane \texttt{qml.AmplitudeDamping} \\
Decoherence Strength & Single shot, probability $p \sim U(0, 0.5)$ (independent for each sample) \\
Random Seed & 42 \\
\noalign{\smallskip}\hline
\end{tabular}
\end{table*}

\begin{table*}[htbp]
\centering
\caption{Parameters for the Regression Task}
\label{tab:regression}
\begin{tabular}{>{\raggedright}p{0.4\linewidth} >{\raggedright\arraybackslash}p{0.5\linewidth}}
\hline\hline\noalign{\smallskip}
\textbf{Parameter} & \textbf{Value / Method} \\
\noalign{\smallskip}\hline\noalign{\smallskip}
Model Type & \texttt{CoherenceNet} (Deep network with residual blocks) \\
Data Scaling & Standardization (\texttt{StandardScaler}) \\
Learning Rate (\texttt{lr}) & Select the optimal value from [1e-3, 1e-4] \\
Weight Decay (\texttt{weight\_decay}) & Select the optimal value from [1e-4, 1e-5] \\
Batch Size & Full-batch \\
Loss Function ($C_{l_1}$) & Mean Absolute Error (MAE) -- \texttt{nn.L1Loss} \\
Loss Function ($C_{g}, C_{cof}$) & Mean Squared Error (MSE) -- \texttt{nn.MSELoss} \\
\noalign{\smallskip}\hline
\end{tabular}
\end{table*}

\begin{table*}[htbp]
\centering
\caption{Parameters for the Classification Task}
\label{tab:classification}
\begin{tabular}{>{\raggedright}p{0.4\linewidth} >{\raggedright\arraybackslash}p{0.5\linewidth}}
\hline\hline\noalign{\smallskip}
\textbf{Parameter} & \textbf{Value / Method} \\
\noalign{\smallskip}\hline\noalign{\smallskip}
Model Type & \texttt{ClassificationCoherenceNet} \\
Architecture & \texttt{CoherenceNet} + Sigmoid output \\
Label Generation & If $C(\rho_t) < 0.8 \cdot C(\rho_0)$, the label is 1, otherwise 0. \\
Threshold Factor ($z$) & 0.8 (Fixed value) \\
Loss Function & Binary Cross-Entropy (BCE) -- \texttt{nn.BCELoss} \\
Other Hyperparameters & Same as the regression task (Adam, hyperparameter search, etc.) \\
\noalign{\smallskip}\hline
\end{tabular}
\end{table*}

\clearpage

\bibliographystyle{IEEEtran}
\bibliography{ref}

\end{document}